\documentclass[aps,prl,twocolumn,showpacs]{revtex4-1}%

\usepackage{amsfonts}
\usepackage{amsmath}
\usepackage{amssymb}
\usepackage{graphicx}
\usepackage{dcolumn}
\usepackage{bm}

\begin{document}
\title{Response to Comment by Annette Bussmann-Holder (arXiv:0909.3603).}
\author{V. G. Kogan, C. Martin, and R. Prozorov}
%\email{kogan@ameslab.gov}
\affiliation{Ames Laboratory and Department of Physics \& Astronomy, 
Iowa State University,
Ames, Iowa 50011}


% 74.20.-z Theories and models of superconducting state
% 74.20.Rp Pairing symmetries (other than s-wave)
% 74.70.Tx Heavy-fermion superconductors

\date{\today}
\maketitle

The author of this Comment \cite{bussmann} (ABH) criticizes severely our paper on \textit{``Superfluid density and specific heat within a self-consistent scheme for a two-band superconductor"}, \prb \textbf{80}, 014507 (2009) \cite{KMR}. ABH states that ``Due to variety of simplifications and unphysical consequences of those, erroneous results and misleading conclusions are obtained". We therefore respond to the critique point by point.

1. ABH states after Eq.(3) of the Comment: ``The assumption made in [1] that $(N_1+N_2)/N(0)=1$ restricts the system to a constant doping level and limits all further applications."

This statement seems to be a misunderstanding. In our text, $N_1$ and $N_2$ are partial densities of states on two bands, so that  $N_1+N_2=N(0)$, the total DOS. This is not an assumption and not a restriction.

2. Another ``unwelcome simplification", according to ABH, is to assume the energy scales $\omega_D$ are the same on both bands.

To this we may say the following: our work do not pretend to be a microscopic first-principles description of two-band superconductors. Our goal was to develop a self-consistent weak-coupling based procedure for evaluation of the $T$ dependent order parameters $\Delta_{1,2}$, the penetration depth, and the specific heat jump, without introducing too many unknown and uncontrollable parameters and using experimentally available information, in order to improve on the widely employed ``alpha-model". The latter uses the BCS temperature dependence for both $\Delta$'s. Introducing different energy cutoffs on two bands would not improve the procedure suggested. Within the weak-coupling model of BCS-Gor'kov-Eilenberger this cutoff is needed as an approximate quantity which nevertheless allows one to relate, e.g., $T_c$ and $\Delta(0)$. Moreover, the weak-coupling theory can be formulated to contain the measurable $T_c$ instead of vaguely defined $\omega_D$. Introducing different, and as a rule unknown, $\omega_D$'s  on two bands would not change by much the results since $\omega_D$'s enter our quantity $S=\ln(2\omega_D/1.76 T_c)$ only under the log-sign. Hence, yes, we use this simplification, but it is welcome because it provides a tight self-consistent description similar to many other models based on the weak-coupling scheme.

3. We do not agree with the further analysis of ABH in which she treats a non-linear Eqs.(4a,4b) for $\Delta_{1,2}$ as a linear homogeneous system requiring a zero determinant, Eqs.(6,7). The point missed in this analysis is that  $F_{1,2}$ (ABH notation) are themselves depend on $\Delta$'s. It is only at $T=T_c$ the system becomes linear, which allows one to express $T_c$ in terms of coupling constants and of partial DOS'. Hence, we find Eqs.(6,7) of ABH incorrect along with the   harsh conclusions that follow.

\end{document}